\documentclass[12pt,preprint]{emulateapj}

\def\be{\begin{equation}}
\def\ee{\end{equation}}
\def\bea{\begin{eqnarray}}
\def\eea{\end{eqnarray}}
\usepackage{graphicx}

\usepackage{color,txfonts}
\begin{document}

\title{A GeV source in the direction of supernova remnant CTB 37B}

\author{Yu-Liang Xin\altaffilmark{1,2}, Yun-Feng Liang\altaffilmark{1,2}, Xiang Li\altaffilmark{1,2}, Qiang Yuan\altaffilmark{3}, Si-Ming Liu\altaffilmark{1},  and Da-Ming Wei\altaffilmark{1}}
\altaffiltext{1}{Key laboratory of Dark Matter and Space Astronomy, Purple Mountain Observatory, Chinese Academy of Sciences, Nanjing 210008, China;}
\altaffiltext{2}{University of Chinese Academy of Sciences, Yuquan Road 19, Beijing, 100049, China;}
\altaffiltext{3}{Department of Astronomy, University of Massachusetts, Amherst, MA 01002, USA.}
\email{liusm@pmo.ac.cn(SML), dmwei@pmo.ac.cn(DMW)}

\begin{abstract}
Supernova remnants (SNRs) are the most attractive candidates for the
acceleration sites of Galactic cosmic rays. We report the detection of 
GeV $\gamma$-ray emission with the Pass 8 events recorded by Fermi Large 
Area Telescope (Fermi-LAT) in the vicinity of the shell type SNR CTB 37B
that is likely associated with the TeV $\gamma-$ray source HESS J1713-381.
The photon spectrum of CTB 37B is consistent with a power-law with an 
index of $1.89\pm0.08$ in the energy range of $0.5-500$ GeV, and the
measured flux connects smoothly with that of HESS J1713-381 at a few
hundred GeV. No significant spatial extension and time variation are 
detected. The multi-wavelength data can be well fitted with either a 
leptonic model or a hadronic one. However, parameters of both models
suggest more efficient particle acceleration than typical SNRs.
Meanwhile, the X-ray and $\gamma$-ray spectral properties of CTB 37B 
show that it is an interesting source bridging young SNRs dominated by 
non-thermal emission and old SNRs interacting with molecular clouds.

\end{abstract}


\setlength{\parindent}{.25in}

\section{Introduction}

Among various suggested scenarios, the leading sources of Galactic cosmic 
rays (CRs) below the spectral knee are believed to be supernova remnants 
\citep[SNRs; see][for a review]{Hillas2005}. The original idea of supernova 
as the source of CRs is motivated by the fact that a reasonable fraction
of the kinetic energy of the supernova ejecta is comparable to that 
needed to sustain the Galactic CRs \citep{Baade1934}. The SNR scenario for 
the origin of CRs, however, was widely accepted only after the development 
of the shock acceleration theory of particles in 1970s' \citep{Drury1983,
Hillas1984,Bell1978a,Bell1978b}. Direct observational evidence was absent 
for a long time, until in recent years several important discoveries were made 
due to the quick development of the $\gamma$-ray detection \citep[e.g.,][]
{Aharonian2006a,Aharonian2008}. Fermi-LAT collaboration 
reported the detection of the characteristic ``$\pi^0$ bump'' from 
sub-GeV $\gamma$-rays of SNRs IC443 and W44, which has been considered as the 
most direct evidence for the presence of relativistic nuclei acceleration 
in SNRs \citep{Fermi2013}\footnote{The hard sub-GeV spectrum of W44 
measured by AGILE was also interpreted due to $\pi^0$-decay 
origin \citep{AGILE2011}.}. Neutral pions produced in proton-proton (more
generally nuclei-nuclei) collisions can decay and give rise to a $\gamma$-ray
spectrum characterized by a peak at the energy of $m_{\pi^0}c^2/2=67.5$ MeV, 
where $m_{\pi^0}$ is the rest mass of the neutral pion \citep{Dermer1986}. 

However, these are relatively older SNRs interacting with
molecular clouds. Younger remnants should accelerate particles more
efficiently with stronger shocks. A number of relative younger SNRs have been 
identified in GeV and/or TeV band, such as Cassiopeia A \citep{Abdo2010,
Albert2007,Acciari2010}, RX J1713-3946 \citep{Abdo2011,Aharonian2007a}, 
RX J0852-4622 \citep{Tanaka2011,Aharonian2007b}, RCW86 \citep{Yuan2014,
Aharonian2009}, Tycho \citep{Giordano2012,Acciari2011}, HESS J1731-347 
\citep{Yang2014,Abramowski2011}, SN 1006 \citep{Acero2010,Araya2012}.
The shocks in these young SNRs are still strong and can accelerate
particles to higher energies efficiently which makes young SNRs
the preferred targets for studying the acceleration of high energy CRs.
CTB 37B is a relatively young (age $\sim5000$ yr) shell-type SNR located 
at the direction of $(l,b)=(348.7^{0}, +0.3^{0})$ and a large distance of 
$\sim$ 13.2 kpc \citep{Tian2012}. The field of CTB 
37B is one of the most active regions in the Galaxy. Radio observations
reveal that this region is rich in star-burst activities, such as
shell-like structures which are probably associated with recent SNRs 
\citep{Kassim1991}, and OH masers \citep{Frail1996}. 
\citet{Tian2012} pointed out that CTB 37B is outside of 
the CO survey and no molecular cloud has been detected in the 
region of CTB 37B.
The X-ray emission of CTB 37B was first detected by Ohashi et al. (1996), 
and then by \citet{Aharonian2008} and \citet{Nakamura2009}. With the Chandra 
observation, \citet{Aharonian2008} identified a point source CXOU J171405.7-381031 
in the radio shell of CTB 37B, which has been identified as a new magnetar 
\citep{Sato2010,Halpern2010}. In the region coincident with the radio
shell, the X-ray emission is found to be thermal \citep{Aharonian2008}.
The Suzaku observation revealed non-thermal X-ray emission to the south
of the radio shell with very hard spectrum \citep{Nakamura2009}.
TeV $\gamma$-ray emission has been detected by the High Energy Stereoscopic
System (HESS) \citep[i.e., HESS J1713-381; see][]{Aharonian2006b,
Aharonian2008}. The spectra and morphologies of both the X-ray and TeV 
$\gamma$-rays suggest either a multi-zone leptonic model or a hadronic model 
is responsible for the $\gamma$-ray emission \citep{Aharonian2008,Nakamura2009}.

The GeV $\gamma$-ray emission is expected to provide a more complete view
of the multi-wavelength characteristics of the source. Enlarging the sample 
of $\gamma$-ray SNRs is very important for understanding the CR acceleration 
and interaction in SNRs. A large sample of SNRs enables one to study the 
radiation mechanism, especially its relation with the SNR evolution and the 
ambient environment in a statistical way \citep{Yuan2012,Dermer2013}. 
\citet{Yuan2012} found an interesting correlation between the $\gamma$-ray 
spectra of SNRs and the environmental gas density: The denser the environment 
is, the softer the $\gamma$-ray spectrum is. It is then suggested that the 
inverse Compton scattering (ICS) of high energy electrons is the dominant 
process for those SNRs located in density cavities and the resulting spectra 
are generally hard. On the other hand, the $\pi^0$ decay emission is responsible 
for the $\gamma$-ray emission of interacting systems between SNRs and molecular 
clouds, resulting in softer spectra. 

In this work, we report the analysis of the GeV $\gamma$-ray emission from the 
direction of CTB 37B, with Fermi Large Area Telescope (Fermi-LAT) Pass 8 data.
In Section 2, the data analysis and results are presented, including the spatial, 
spectral and timing analyses. The discussion about the origin of the 
non-thermal radiation based on multi-wavelength spectral energy distribution 
(SED) is given in Section 3. We conclude our work in Section 4.

\section{Data analysis}

\subsection{Data reduction}

\begin{figure*}[!htb]
\centering
\includegraphics[width=\columnwidth]{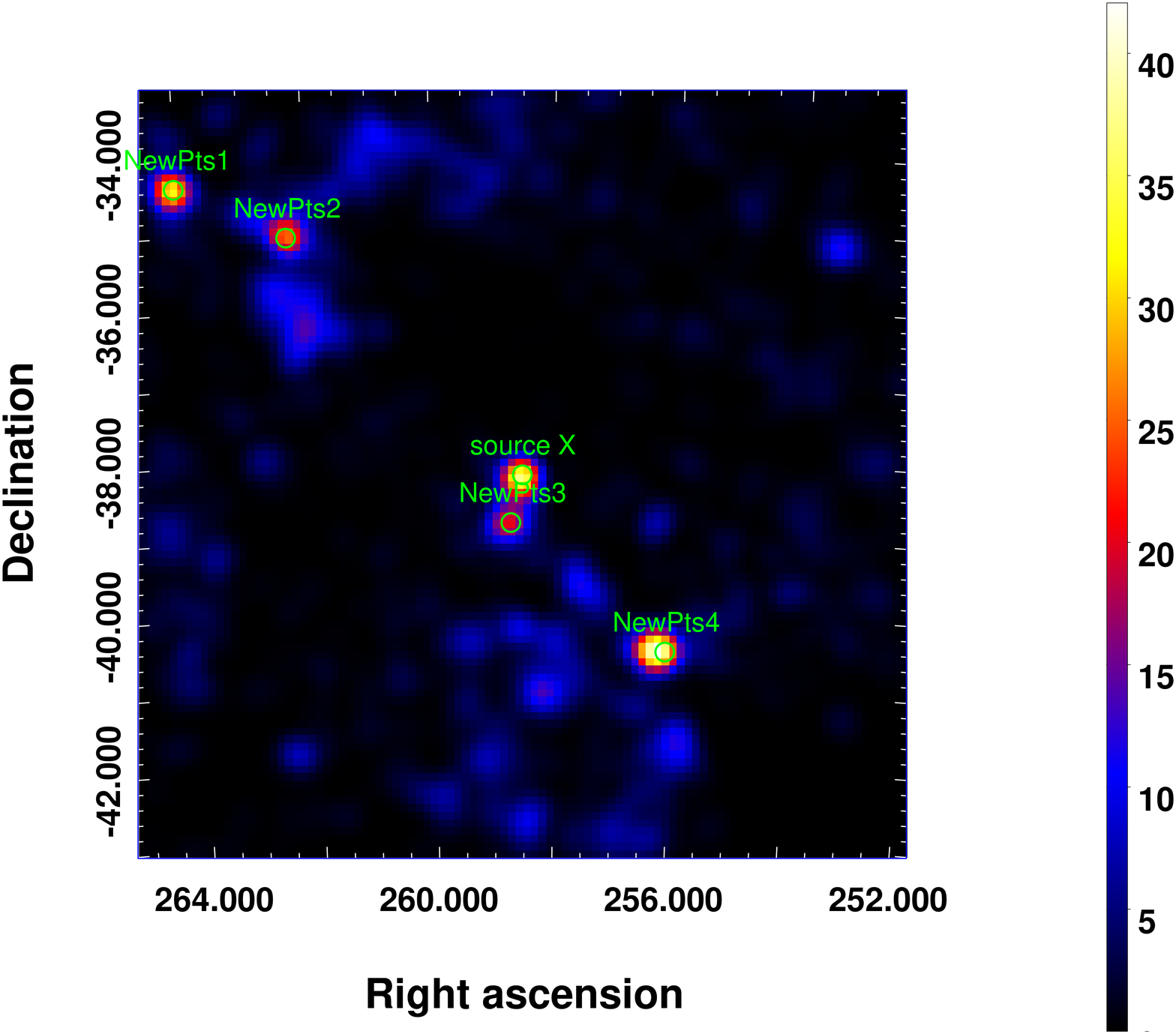}
\includegraphics[width=\columnwidth]{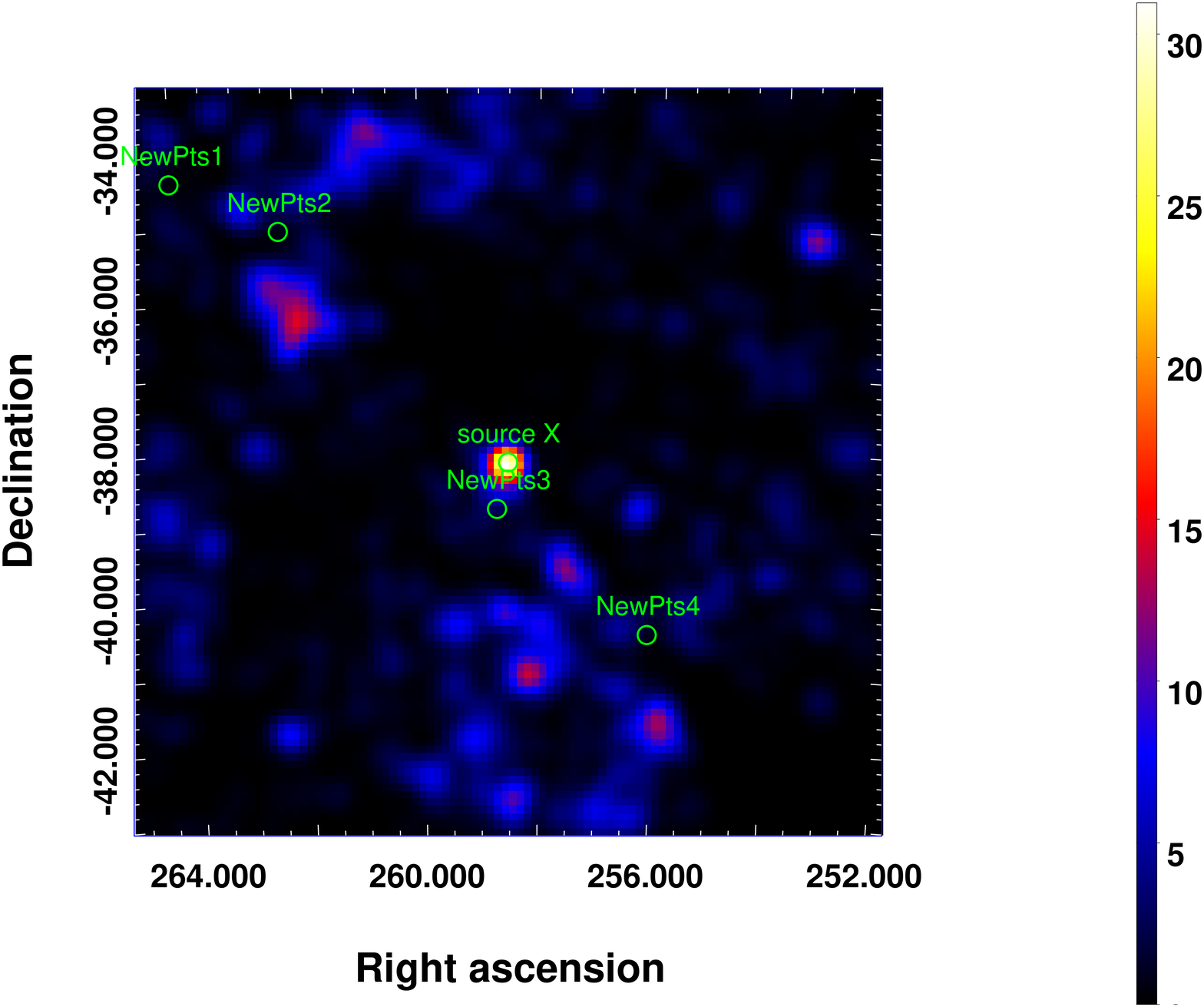}
\caption{$3-500$ GeV TS maps of $10^{\circ} \times 10^{\circ}$ region 
centered at CTB 37B. The left panel is for the model with only the 3FGL 
sources and the diffuse backgrounds. Several bright sources which are not
included in 3FGL are marked with green circles. The right panel is for 
the model with the additional four new sources (except source X, presumably
CTB 37B) as listed in Table \ref{table:newpts}. The maps are smoothed 
with a Gaussian kernel of $\sigma$ = $0.3^\circ$.}
\label{fig:tsmap}
\end{figure*} 

In this analysis, the latest Pass 8 version of the Fermi-LAT data were 
used\footnote{http://fermi.gsfc.nasa.gov/ssc/data}. The data were collected 
from October 27, 2008 (Mission Elapsed Time 246823875) to June 18, 2015 
(Mission Elapsed Time 456279605). The energies of events are cut between 
500 MeV and 500 GeV to avoid the too large point spread function (PSF) 
in lower energy band. We select the ``source'' event class (evclass=128 
\& evtype=3). The maximum zenith angle is adopted to be $90^\circ$ to
minimize the contamination from the Earth limb. We apply a set of quality 
cuts recommended by the LAT team, with {\tt(DATA\_QUAL$>$0) \&\& 
(LAT\_CONFIG==1)}. The analysis is performed in a $14^\circ \times 14^\circ$ 
rectangle region of interest (ROI) centered at the position of CTB 37B
\citep[R.A.$=17^{h}13^{m}58^{s}$, Dec.$=-38^\circ12^{'}00^{''}$;][]{Green2014}. 
The standard LAT analysis software, {\it ScienceTools} version 
{\tt v10r0p5}\footnote{http://fermi.gsfc.nasa.gov/ssc/data/analysis/software/}, 
available from the Fermi Science Support Center, and the instrumental response 
function (IRF) ``P8R2{\_}SOURCE{\_}V6'' are adopted. We use the binned 
likelihood analysis method with {\tt gtlike} to fit the data. The Galactic 
and isotropic diffuse background models used are {\tt gll\_iem\_v06.fits} 
and {\tt iso\_P8R2\_SOURCE\_V6\_v06.txt}\footnote
{http://fermi.gsfc.nasa.gov/ssc/data/access/lat/BackgroundModels.html}.
The point sources in the third Fermi catalog \citep[3FGL;][]{Acero2015}
are included in the model, generated with the user-contributed software 
{\tt make3FGLxml.py}\footnote{http://fermi.gsfc.nasa.gov/ssc/data/analysis/user/}.

\subsection{Source detection}

During the likelihood fittings, the normalizations and spectral parameters 
of the sources within $7^\circ$ around CTB 37B, together with the 
normalizations of the two diffuse backgrounds, are left free. Firstly, 
we performed the fitting with the 3FGL sources and the diffuse backgrounds.
A $10^{\circ}\times10^{\circ}$ Test Statistic (TS) map for photons above
3 GeV is created with {\tt gttsmap}, as shown in the left panel of
Fig. \ref{fig:tsmap}. The TS map shows that there are extra sources
beyond the 3FGL catalog. We mark five bright new sources with green
circles in this plot. Especially, we find that at the position of
CTB 37B, there is an evident excess with peak TS value of $\sim 79$,
which is marked as source X. Then we add these sources in the model,
assuming power-law spectra and the approximate locations read from the 
TS map, and fit the data again. The precise positions are obtained using
{\tt gtfindsrc} command. The TS map with the additional four sources
(excluding source X) included in the model can be seen in the right
panel of Fig. \ref{fig:tsmap}, which looks much smoother. 
The fitting positions and TS values of these new sources are listed 
in Table \ref{table:newpts}. 

\begin{center}
\begin{figure*}[!htb]
\centering
\includegraphics[width=0.8\textwidth]{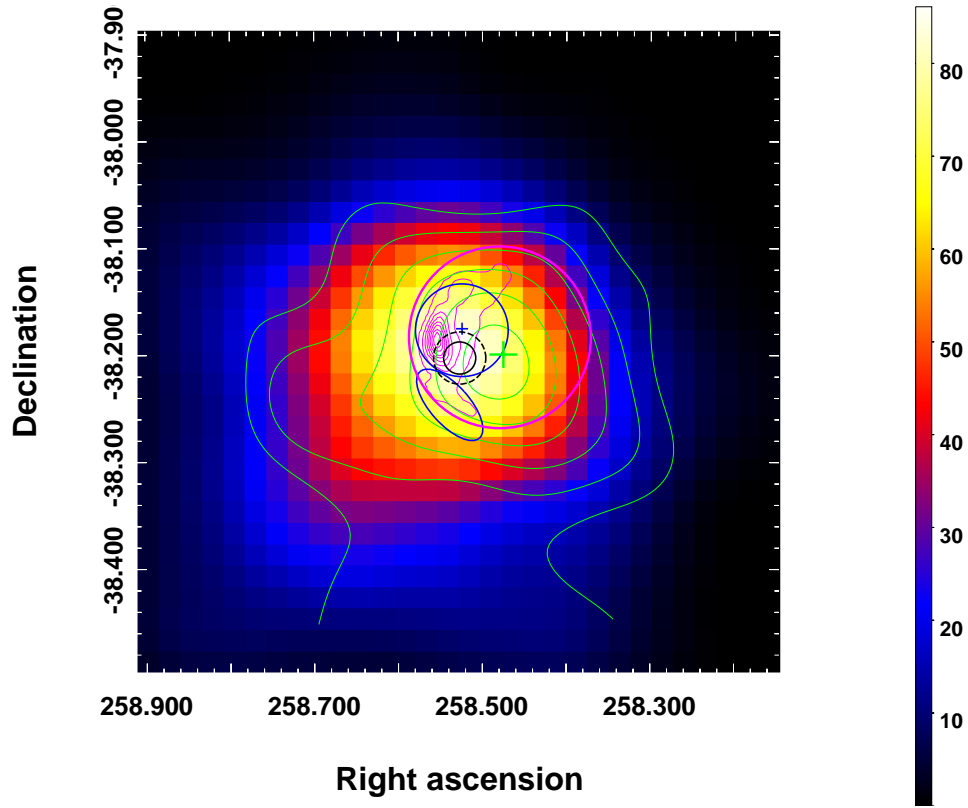}
\caption{Zoom-in of the TS map, for a region of $0.6^{\circ}\times0.6^{\circ}$
centered at the best-fitting position of source X. The image was created 
using a grid of $0.02^\circ$ and smoothed with a $\sigma=0.3^\circ$ Gaussian 
kernel. Black circles show the $1\sigma$ (solid) and $2\sigma$ (dash) error 
circles of the fitting position of source X. Magenta contours represent the 
radio image of CTB 37B at 843 MHz from SUMSS \citep{Mauch2003}. The radio 
size of the SNR CTB 37B is indicated by the magenta circle with a radius of 
$5.1'$. Green contours show the HESS image of TeV $\gamma$-ray emission 
with the green plus sign marking the peak position \citep{Aharonian2008}. 
The location of the magnetar CXOU J171405.7-381031 is marked by the blue 
plus sign. The blue circle and ellipse represent X-ray emission regions 1 
and 2 of Fig 1 in \citet{Nakamura2009}.}
\label{fig:tsmap2}
\end{figure*}
\end{center}

We focus on source X in the following analysis. The TS value of source
X is found to be 102.6, which corresponds to a significance of 
$\sim9.5\sigma$ for 4 degrees of freedom (dof). 
Using {\tt gtfindsrc} tool, we find the best-fitting position of 
source X is R.A.$=258.527^\circ$, 
Dec.$=-38.2027^\circ$ with $1\sigma$ uncertainty of $0.015^\circ$. 
To better see the spatial relation of source X and SNR CTB 37B in other
wavelengths, we show a zoom-in of the TS map for a region of $0.6^{\circ} 
\times 0.6^{\circ}$ centered on source X in Fig. \ref{fig:tsmap2}.
The radio contours at 843 MHz from the Sydney University Molonglo Sky
Survey \citep[SUMSS;][]{Mauch2003}, and the TeV $\gamma$-ray contours 
from HESS J1713-381 \citep{Aharonian2008} are overplotted. The position
of the X-ray point source CXOU J171405.7-381031
\citep[R.A.$=17^{h}14^{m}5.758^{s}$, Dec.$=-38^\circ10^{'}31.32^{''}$;]
[]{Aharonian2008}, which was identified
as a magnetar \citep{Nakamura2009}, is marked by a blue plus. The blue
circle and ellipse show the regions 1 and 2 of \citet{Nakamura2009},
which show thermal and non-thermal diffuse emissions, respectively.
As can be seen, the position of source X is in good coincidence with
the radio and TeV $\gamma$-ray images of CTB 37B. CXOU J171405.7-381031 
is located slightly to the north of the best-fitting position of source
X (outside the $2\sigma$ error circles). As discussed in 
\citet{Aharonian2008}, the absence of associated extended non-thermal 
emission around CXOU J171405.7-381031 argues strongly against the pulsar 
wind nebula (PWN) origin of the TeV $\gamma$-ray emission. If source
X is associated with CXOU J171405.7-381031, its $\gamma$-ray emission
(from the pulsar) should be point-like, and then one may have difficulty
to explain the displacement of their positions. Therefore source X 
is more likely the counterpart of SNR CTB 37B. 

\begin{table}[!htb]
\centering
\caption {Coordinates and TS values of the five newly added point sources including source X}
\begin{tabular}{cccc}
\hline \hline
Name & R.A. [deg] & Dec. [deg] & TS \\
\hline
source X & $258.527$ & $-38.2027$ & $102.6$\\
NewPts1  & $264.032$ & $-34.3641$ & $74.2$ \\
NewPts2  & $262.291$ & $-35.0589$ & $61.2$ \\
NewPts3  & $258.718$ & $-38.8246$ & $73.8$ \\
NewPts4  & $256.209$ & $-40.4863$ & $94.4$ \\
\hline
\hline
\end{tabular}
\label{table:newpts}
\end{table}

\subsection{Spatial extension}
The radio diameter of CTB 37B is about $10'$, as shown by the magenta
circle in Fig. \ref{fig:tsmap2}. Such a size may be too small to be 
resolved with Fermi-LAT. As a test, we use the SUMSS radio image, the 
HESS TeV $\gamma$-ray image, as well as uniform disks centered at the
best-fitting position with different radii as spatial templates and 
re-do the fittings. The TS values for the SUMSS radio and HESS TeV 
$\gamma$-ray templates are 102.4 and 97.5, respectively. For uniform 
disks with different radii, the TS values range between 100 and 105
and the upper limit of the GeV emission radius corresponding 
to $1\sigma$ confidence level is given as $7^{'}$.
The extended spatial templates do not improve the significance of CTB 
37B significantly. In the following SED analysis, we will keep the point 
source assumption.

\subsection{Spectral analysis}

\begin{figure}[!htb]
\centering
\includegraphics[angle=0,scale=0.35,width=0.5\textwidth,height=0.3\textheight]{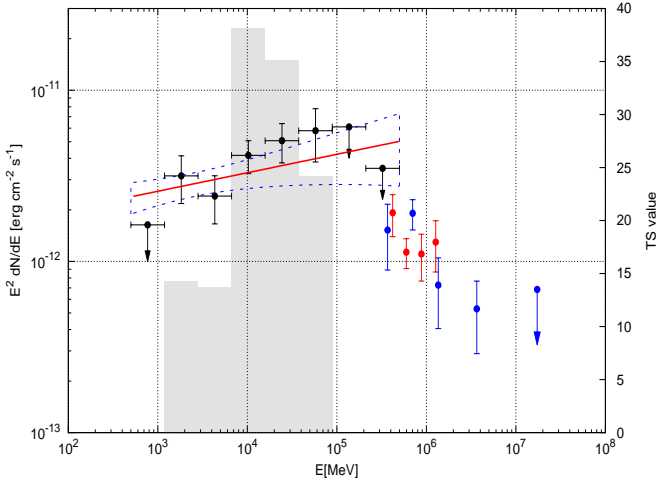}%
\hfill
\caption{SED of source X. The results of Fermi-LAT data are shown by 
black dots, with arrows indicating the $95\%$ upper limits. The red 
solid line is the best-fitting power-law in the energy range of 0.5
to 500 GeV, and the blue dashed butterfly shows the $68\%$ range
of the global fitting. The gray histogram denotes the TS value 
for each energy bin. High energy data points (red and blue) are 
from HESS observations \citep{Aharonian2006b,Aharonian2008}.} 
\label{fig:sed}
\end{figure}

The power-law index of source X is found to be $1.89\pm0.08$ in the
energy range of $0.5-500$ GeV and the integral photon flux is
$(3.33\pm0.63)\times10^{-9}$ photon cm$^{-2}$ s$^{-1}$. The $\gamma$-ray
luminosity between 500 MeV and 500 GeV is $5.13\times 10^{35}\, 
(d/13.2\ {\rm kpc})^2$ erg~s$^{-1}$, where a distance $d = 13.2$ kpc
\citep{Tian2012} is adopted.

To derive the SED of source X at different energies, we bin the data 
with 8 equal logarithmic energy bins between 500 MeV and 500 GeV, and perform 
the same likelihood fitting analysis to the data. The flux normalizations 
of all the sources  within $7^\circ$ of source X are left free, 
while the spectral indices are fixed. For source X, both the normalization 
and spectral index are left free. The remaining free parameters include 
the normalizations of the two diffuse backgrounds. The fitting results are 
shown in Fig. \ref{fig:sed}. Note that for the energy bins with TS value of 
sources X smaller than 5, we give the upper limits at $95\%$ confidence
level, shown by the black arrows in Fig. \ref{fig:sed}. We find that
the Fermi-LAT data connect with the TeV $\gamma$-ray SED of HESS J1713-381
smoothly at a few hundred GeV suggesting that source X is the GeV 
counterpart of SNR CTB 37B.

\subsection{Timing analysis}

\begin{figure}[!htb]
\centering
\includegraphics[angle=0,scale=0.35,width=0.5\textwidth,height=0.3\textheight]{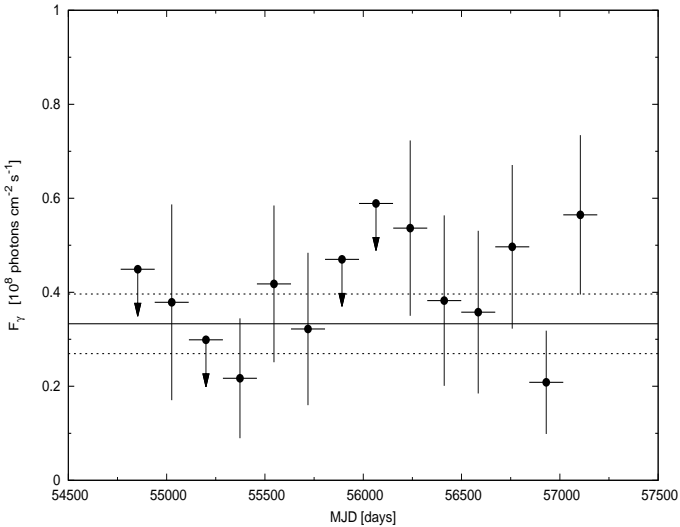}%
\hfill
\caption{Light-curve of source X. The horizontal solid and dashed lines 
are average flux and its $1\sigma$ range from the whole data set.}
\label{fig:lt}
\end{figure}

The emission from SNR is expected to be stable at the time scale of years.
As a check, we perform the time analysis of source X. The near 7 years'
data are binned into 14 time bins equally. The fitting 
method is the same as the SED analysis. The results are shown in Fig. 
\ref{fig:lt}. The arrows indicate the 95\% upper limits of those time 
bins, whose TS values are smaller than 5. No obvious long term variation 
is found, which is consistence with the emission expected from an SNR.

\section{Discussion}

The spatial and spectral association between source X and
the HESS observation of SNR CTB 37B suggests source X being the
GeV counterpart of this SNR. The $\gamma$-ray emission can be either from
the ICS of high energy electrons or the $\pi^0$ decay due to the inelastic
$pp$ collisions. Due to the lack of non-thermal X-ray
emission in the same region of the TeV $\gamma$-ray emission, 
\citet{Aharonian2008} argued that the TeV $\gamma$-rays may have a
hadronic origin. The Suzaku observations reveal, however, both thermal
and non-thermal diffuse X-ray emissions from the western part of the
radio shell, which suggests alternatively a multi-zone leptonic scenario
for the $\gamma$-ray emission \citep{Nakamura2009}. We discuss both the
leptonic and hadronic models in light of the multi-wavelength data,
including the Fermi-LAT ones.

In the modeling, both the spectra of electrons and protons are assumed to
be power-laws with exponential cutoffs, $dN/dE_i \propto E_i^{-\alpha_i} 
\exp(-E_i/E_{i,\rm cut})$, where $i = e$ or $p$, $\alpha_i$ and 
$E_{i, \rm cut}$ are the spectral index and the cutoff energy, respectively. 
The distance of CTB 37B is adopted to be 13.2 kpc \citep{Tian2012}, and 
the radius is taken to be $r\approx20$ pc which corresponds to an angular 
size of $5.1'$ at such a distance. The radiation field includes the cosmic
microwave background (CMB), an infrared field with $T=30$ K and energy
density $u=1$ eV cm$^{-3}$, and an optical field with $T=6000$ K and
$u=1$ eV cm$^{-3}$ \citep{Porter2006}. The gas density is adopted to 
be $0.5$ hydrogen cm$^{-3}$, as inferred from the X-ray observations 
\citep{Aharonian2008,Nakamura2009}.

For the region which is coincident with the SNR shell (region 1), no 
non-thermal emission has been detected \citep{Aharonian2008,Nakamura2009}. 
The $1-5$ keV flux of the thermal emission is $\sim3\times10^{-13}$
erg cm$^{-2}$ s$^{-1}$ \citep{Aharonian2008}. The $2-10$ keV unabsorbed
flux of the thermal X-ray emission from region 1 is estimated to be 
$\sim6\times10^{-13}$ erg cm$^{-2}$ s$^{-1}$, based on the non-equilibrium
ionization model employed in \citet{Aharonian2008}. This thermal flux is
adopted as an upper limit of the non-thermal emission from the SNR.
 
\subsection{Leptonic model}

\begin{table*}
\centering
\caption {Parameters for the models}
\begin{tabular}{cccccccccc}
\hline \hline
Model & $\alpha_p$ & $\alpha_e$ & $E_{p,\rm cut}$ & $E_{e, \rm cut}$ & $W_p$ & $W_e$ & $B$ & $n_{\rm gas}$\\
      &            &            & (TeV)     & (TeV)     & ($10^{50}$ erg) & ($10^{50}$ erg) & ($\mu$G) & (cm$^{-3}$)\\
\hline
leptonic  & $-$    & $1.65$ & $-$   & $0.92$  & $-$   & $0.09$ & $100$ & $0.5$ \\
\hline
hadronic  & $1.86$ & $1.65$ & $3.0$ & $0.65$ & $50$  & $0.03$ & $200$ & $0.5$ \\
\hline
\hline
\end{tabular}
\label{table:model}
\tablecomments{The total energy of relativistic particles, $W_{e,p}$, is 
calculated for $E > 1$ GeV.}
\end{table*}

\begin{figure*}[!htb]
\centering
\includegraphics[width=\columnwidth]{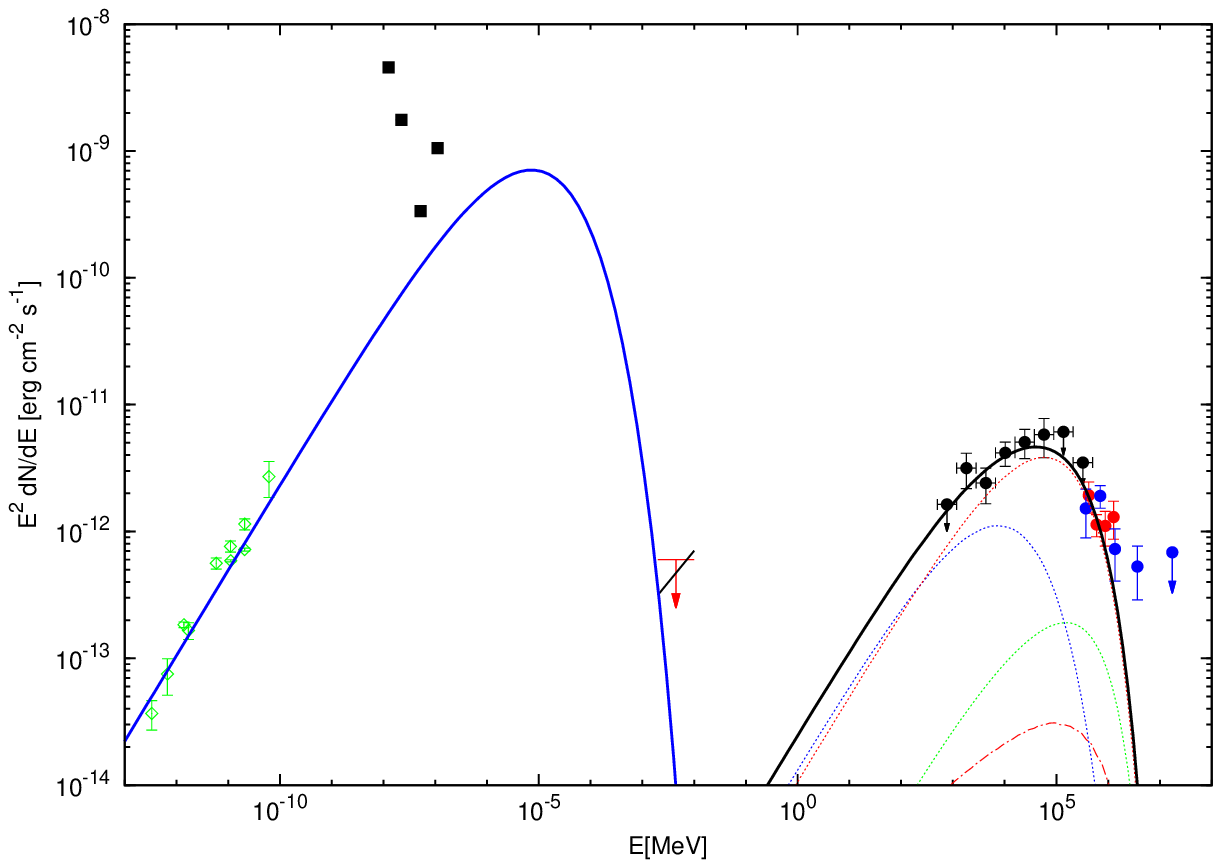}
\includegraphics[width=\columnwidth]{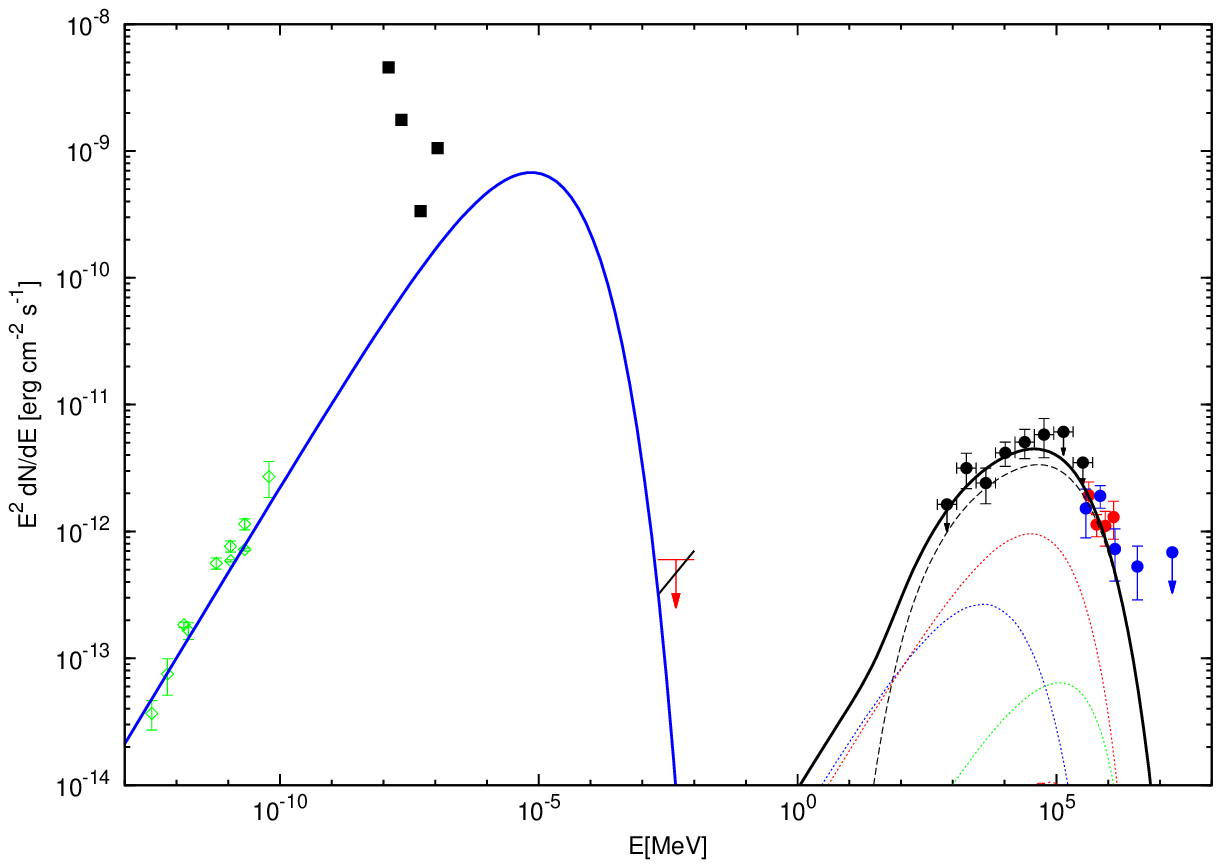}
\caption{Multi-band SED of CTB 37B. Radio data are from \citet{Kassim1991}.
Infrared data marked by black squares are taken from the Infrared Astronomical
Satellite (IRAS) survey. Red arrow shows the thermal X-ray emission from 
region 1 \citep{Aharonian2008}, and the black slash shows the non-thermal 
X-ray emission from region 2 \citep{Nakamura2009}. The left panel is for the 
leptonic model, and the right panel is for the hadronic model. For the $\gamma$-ray 
emission, the contributions from ICS, bremsstrahlung, and $\pi^0$-decay are shown
by dotted, dot-dashed, and dashed lines, respectively. The three components
of ICS $\gamma$-ray emissions from scatterings off CMB (blue), infrared 
(red), and optical (green) photon fields are shown separately.} 
\label{fig:multi-sed}
\end{figure*} 

In the leptonic model, the radio to X-ray emission is due to synchrotron
radiation of relativistic electrons in the magnetic field, and the 
$\gamma$-ray emission is produced through ICS in the background radiation
field and bremsstrahlung in the interstellar medium. To fit the radio to
TeV $\gamma$-ray data, we find $\alpha_{\rm e} \approx 1.65$, $E_{e, \rm cut} 
\approx 0.92~{\rm TeV}$, a magnetic field $B\approx100$ $\mu$G, and a total
electron energy above 1 GeV $W_{\rm e}\approx 9.0 \times 10^{48}~{\rm erg}$.
The fitting SED is shown in the left panel of Fig. \ref{fig:multi-sed}. 
The model parameters are compiled in Table \ref{table:model}. 

We may compare the model parameters with those from several other young SNRs
which are primarily thought to be leptonic sources, such as RX J1713-3946 
\citep{Abdo2011,Yuan2011}, RX J0852-4622 \citep{Tanaka2011}, RCW86 
\citep{Yuan2014}, HESS J1731-347 \citep{Yang2014}, and SN 1006 
\citep{Acero2010,Araya2012}. The cutoff energies of electrons for these 
sources are typically of several tens TeV, significantly higher than the 
value of $\sim$TeV for CTB 37B. This could be partially due to that the
magnetic field is relatively higher for this source. For $B\approx100$
$\mu$G, electrons with energies higher than $\sim$ TeV will cool
down if they are accelerated in the early stage of the SNR whose age
was estimated to be a few thousand years \citep{Aharonian2008}. 
This age estimation was also supported by the magnetar 
characteristic age \citep[$\sim10^{3}$ yrs;][]{Sato2010}.
The total energy of electrons is also higher than that of other SNRs, 
which are of the order of $10^{47}-10^{48}$ erg \citep{Yang2014}. 
Note that this energy estimate should suffer from the uncertainty of 
the distance estimate.

We note that the X-ray emission of CTB 37B is dominated by 
a thermal component and the weak non-thermal component has a very hard 
spectrum.  These characteristics are distinct from that of the other 
young SNRs whose X-ray emission has a prominent non-thermal soft component 
produced by TeV electrons via the synchrotron process \citep{Takahashi2008,
Slane2000,Bamba2012}. The SNR RCW 86 is more similar to CTB 37B 
than the others in the sense that it also has a prominent thermal X-ray 
component \citep{Bamba2000,Borkowski2001}. While the model parameters of
CTB 37B are also extreme compared with that of leptonic model for RCW 86
\citep{Yuan2014}.


\subsection{Hadronic model}

The right panel of Fig. \ref{fig:multi-sed} shows the results of the
hadronic model fitting. The model parameters are given in Table 
\ref{table:model}. The proton spectral index is found to be about 
$1.9$, and the cutoff energy is about 3 TeV. The magentic field
is even stronger to suppress the ICS contribution from the electrons.
While the spectral index does not differ much from the expectation
of diffusive shock acceleration, the maximum energy seems to be too
low for typical SNR shocks \citep{Gaisser1990}. The total energy of 
relativistic protons above 1 GeV is $W_p \approx 5 \times 10^{51} 
(n/0.5\,\mathrm{cm}^{-3})^{-1}\ \mathrm{erg}$. For typical kinetic
energy released by a core-collapse supernova, $E_k\sim10^{51}$ erg,
such an energy of the CR particles seems to be too high. 
One possibility is that the progenitor of CTB 37B is an extremely
massive star which results in a hypernova explosion with a much
higher energy realease than typical supernova. 

On the other hand, it is possible that there are some high-density shocked 
clouds which do not emit thermal X-rays owing to the low-temperature in 
the post-shock region \citep{Inoue2012}. The actual density of gas 
interacting with CRs could be much higher than that inferred from the 
thermal X-ray emission. Assume a density of $\sim 5$ cm$^{-3}$, the 
required energy of protons reduces to be $\sim 5\times10^{50}\ \mathrm{erg}$.

The synchrotron power spectrum peaks near 
$\sim 19\, (E_{\rm e}/1\, {\rm TeV})^2\, (B/1\, {\rm Gauss})$ keV.
To produce X-ray emission by sub-TeV electrons via the synchrotron process, 
the magnetic field strength should exceed 0.1Gauss which is too high to 
account for the non-thermal diffusive X-ray emission in region 2. We therefore 
suggest that this diffusive non-thermal component is likely associated with 
a different energetic electron population with energy in the TeV range.

\section{Conclusion}

In this work we analyze the $\sim7$ years $\gamma$-ray data from Fermi-LAT 
in the field of SNR CTB 37B. A point-like source with a significance of
$\sim9.5\sigma$ has been detected, with position coincident with the
radio and TeV $\gamma$-ray images of CTB 37B. The spectral index in
$0.5-500$ GeV range is found to be $1.89\pm0.08$, and the SED matches
well with the HESS observations at a few hundred GeV energies. We do
not find significant spatial extension and variability of the source,
which are also consistent with the expectation from CTB 37B. This GeV 
source is suggested to be the GeV counterpart of SNR CTB 37B.

The multi-wavelength data can be well fitted by a leptonic or a 
hadronic model. However, the model parameters of both scenarios seem
to be extreme compared with other similar SNRs. The estimated total 
energies of relativistic particles are too high for both scenarios, 
which might be due to an over-estimate of the distance of the source. 
The cutoff energy is found to be $\sim$TeV, which is much lower than 
that in other young SNRs mentioned in the previous section.

While the $\gamma$-ray spectrum of CTB 37B is softer than 
the other young SNRs, its GeV spectrum is harder than that of older SNRs, 
such as IC443 and W44. Only thermal X-ray emission has been detected from 
older SNRs \citep{Yamaguchi2009,Uchida2012}. The X-ray and $\gamma$-ray 
spectral properties suggest that it is an interesting source bridging 
young SNRs dominated by non-thermal emission and old SNRs interacting 
with molecular clouds.

\section*{Acknowledgments}
This work was supported in part by 973 Programme of China under grants 2013CB837000 and 2014CB845800, National Natural Science of China under grants 11361140349, 11273063, 11433009, 11173064, 11233001 and 11233008, the Foundation for Distinguished Young Scholars of Jiangsu Province, China (No. BK2012047) and the Strategic Priority Research Program (Grant No. XDB09000000).

\end{document}